# Positive-tone Nanolithography of Antimony Trisulfide with Femtosecond Laser Wet-etching

**Abhrodeep Dey[1,2], Uwe Hübner[2], Albane Benardais[3], Xiaofei Wu[2], Andrea Dellith[2], Jan Dellith[2], Torsten Wieduwilt[2], Henrik Schneidewind[2], Markus A Schmidt[2,4,5], Virginie Nazabal[3], Volker Deckert[1,2,4,6], Jer-Shing Huang[1,2,7,8], Wei Wang[1,2*]**

[1] Institute of Physical Chemistry, Friedrich Schiller University Jena (FSU), 07743 Jena, Germany.

[2] Leibniz Institute of Photonic Technology (IPHT), 07745 Jena, Germany.

[3] Institut des Sciences Chimiques de Rennes (ISCR) – UMR 6226, CNRS, Univ. Rennes, Rennes, F-35000, France.

[4] Abbe Center of Photonics and Faculty of Physics, FSU, 07745 Jena, Germany.

[5] Otto Schott Institute of Material Research, FSU, 07745 Jena, Germany.

[6] Jena Center for Soft Matter (JCSM), FSU, 07743 Jena, Germany.

[7] Research Center for Applied Sciences, Academia Sinica, 11529 Taipei, Taiwan.

[8] Department of Electrophysics, National Yang Ming Chiao Tung University, 30010 Hsinchu, Taiwan.

**Abstract:** Antimony trisulfide ($Sb_2S_3$), as an emerging material for integrated photonic devices, has attracted significant attention due to its high index, low loss, and phase-changing property in the optical regime. However, conventional lithography-based fabrication methods involve complex, time-consuming, multistep processes, rendering the photonic application of $Sb_2S_3$ challenging. Here, we demonstrate that positive-tone fabrication of $Sb_2S_3$ nanostructures using wet-etch femtosecond laser processing, a straightforward technique for the engraving of micro- and nanoscale structures, can address major fabrication challenges. The patterning mechanism and factors influencing resolution of $Sb_2S_3$ thin film structures deposited on quartz (transmissive) and gold (reflective) substrates are experimentally investigated and supported by theoretical modelling. Using this approach, the smallest linewidth fabricated is measured at 178 nm. Consequently, multiple test patterns are demonstrated showing versatile functionalities. Functional Fresnel Zone Plates (FZPs) with varying focal length are fabricated and characterized. This study provides a significantly simplified approach for realizing $Sb_2S_3$ based integrated photonic devices.

**1. Introduction:**

In recent years, $Sb_2S_3$ has attracted considerable interest as a potential material for use in lab-on-chip devices operating within the VIS-IR region. The factors that have led to this increased interest are its distinctive optical properties, including high refractive index, low optical loss, phase-changing, high nonlinearity, semiconducting nature, and topological effects[1-3]. Compared to commonly used optical materials such as quartz ($SiO_2$), PMMA, SU-8, and IP-Dip (Nanoscribe GmbH), $Sb_2S_3$ offers a significantly higher refractive index contrast while maintaining low optical losses which is essential for fabricating efficient and compact optical waveguides[4], hollow cavity metalenses and photonic crystals[5,6]. These characteristics make $Sb_2S_3$ an ideal material for creating high-performance diffractive optical elements (DOEs) and metasurfaces[7]. Moreover, $Sb_2S_3$ is a phase-change material[2], which renders it particularly suitable for rewritable photonic devices. It exhibits a high refractive index contrast ($\Delta n \approx 0.6$ at 1550 nm) between its amorphous (a-$Sb_2S_3$) and crystalline (c-$Sb_2S_3$) states[8,9]. The material can undergo multiple phase transitions, switching from amorphous to crystalline through thermal annealing and back to amorphous using high-intensity femtosecond (fs-) laser pulses[2,9]. Recent research has focused on various lithography methods to fabricate $Sb_2S_3$-based nanostructures for optoelectronic devices. H. Liu et al adopted 780 nm fs-laser to fabricated rapid-switching and color-changing devices based on c-$Sb_2S_3$ films[9]. Rui Chen et al. used electron beam lithography (EBL) to integrate $Sb_2S_3$ into a silicon photonic platform, achieving 5-bit operation with high cyclability and low loss[10]. Wang et al. employed direct laser interference patterning to create 2D grating structures on c-$Sb_2S_3$, achieving a power conversion efficiency of 73% in structured solar cells[11]. In subsequent work, Wang et al. demonstrated the versatility of direct patterning $Sb_2S_3$ nanostructures using an Sb-butyldithiocarbamic acid (Sb-BDCA) molecular precursor solution via techniques such as ultraviolet photolithography (UVL), EBL, two-photon absorption lithography (TPAL), and thermal scanning probe lithography[12]. More recently, W. Wang et al. showed that $Sb_2S_3$ nanostructures with customizable lateral and vertical profiles (3D forms) can be achieved by adjusting the electron beam irradiation dose during grayscale EBL[7].

Until now, these patterning approaches have focused on the negative-tone (additive manufacturing) fabrication of $Sb_2S_3$. However, positive-tone (subtractive manufacturing) patterning of $Sb_2S_3$ remains unexplored. In this work, we report the positive-tone fabrication of a-$Sb_2S_3$ structures using a fs-laser assisted wet-etching approach for the first time. Compared to methods such as UVL, EBL, and TPAL, etching-assisted fs-laser processing offers advantages in terms of straightforward, convenience, rapid prototyping, and cost-effective generation of complex patterns and freeform structures[13,14]. UV photolithography offers high throughput and sub-micron resolution for mass production, but is limited to planar fabrication and requires expensive photomasks EBL achieves sub-10 nm resolution, making it ideal for nanoscale structures, while its slow speed, high cost, and lack of scalability limit it to prototyping. TPAL offers sub-micron resolution (~100-200 nm) and smooth 3D patterning, but remains a serial process with scalability dependent on master replication. Compared to these methods, fs-laser wet etching balances precision and material versatility, making it a promising approach for next-generation planar optics. We demonstrate fs-laser assisted wet-etch patterned $Sb_2S_3$ structures on both quartz ($SiO_2$) and gold (Au) substrates, showing the versatility in the fabrication of transmission and reflection based optical devices such as dielectric-loaded waveguides[14], photonic integrated circuits[15], and metasurfaces[16]. Moreover, our patterning capabilities on metals, provide the opportunity to fabricate tunable high-index dielectric loaded plasmonic metasurfaces[17], ultrathin non-volatile displays[9] and plasmon-based spectrometer-free index sensors[18] without the need for an intermediate adhesion layer.

**2. Results and Discussion**

**2.1 Femtosecond laser assisted pattering of a-$Sb_2S_3$**

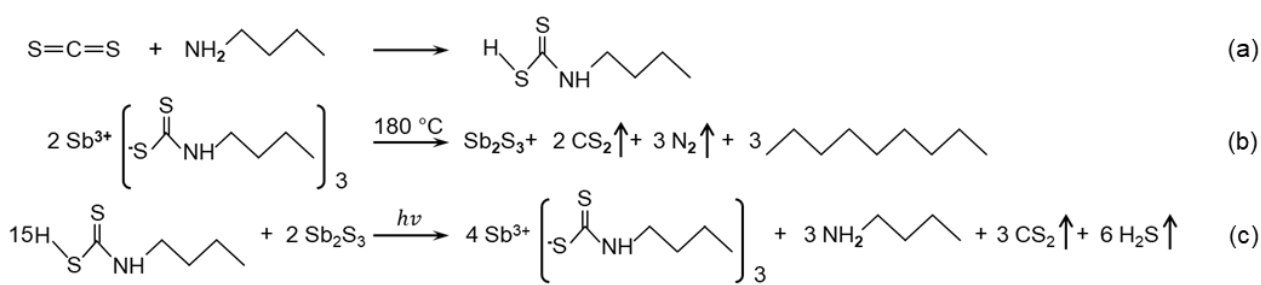

**Scheme 1.** (a) Synthesis of butyldithiocarbamic acid (BDCA) via the reaction of $CS_2$ and n-butylamine. (b) Formation of a-$Sb_2S_3$ via the decomposition of Sb-BDCA complex. (c) Wet-etching of a-$Sb_2S_3$ using a fs-laser pulse in the presence of BDCA.

The foundation of this work involves a BDCA based organic etchant. It is synthesized by the reaction of carbon disulfide ($CS_2$) with n-butylamine (Scheme 1a) and was introduced by Nomura et al. in 1988[19]. BDCA solution's relatively low cost and low toxicity has made it a precursor solvent for spin-coating process in recent years[11,12] while also serving as a versatile solvent for dissolving metal oxides ($M_xO_y$) and hydroxides ($M(OH)_x$). This occurs through the reaction of BDCA with $M_xO_y$ or $M(OH)_x$, yielding metal-organic complexes ($M(BDCA)_x$). Under external stimuli, such as thermal energy or electron beam irradiation, these complexes decompose into corresponding metal chalcogenides (Scheme 1b)[11,12,20].

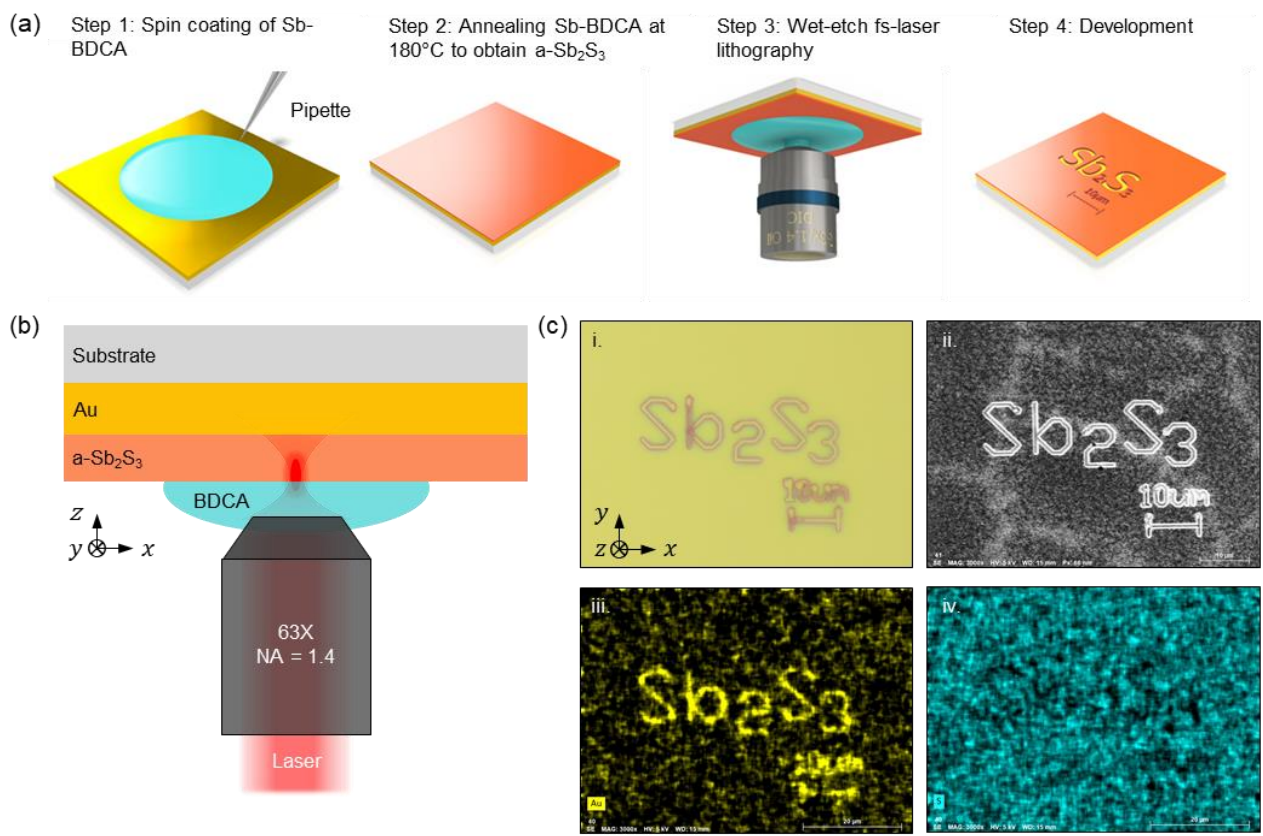

**Figure 1. Schematic of femtosecond laser assisted wet-etching of a-Sb$_2$S$_3$ thin film on gold substrate.** Similar steps are followed for printing structures on quartz substrate. (a) Stepwise display of the structuring procedure. Step 1: Deposition of Sb-BDCA complex onto the target substrate via spin-coating. Step 2: a-Sb$_2$S$_3$ film obtained after annealing the Sb-BDCA coated substrate at 180°C for 10 minutes. Step 3: The desired structures are fabricated using femtosecond laser assisted wet-etch in BDCA acid. Step 4: Excess BDCA and removed material is removed during the development process. (b) Schematic diagram of femtosecond laser assisted wet-etching on a-Sb$_2$S$_3$ using BDCA. (c) i. Bright field (BF) optical microscope image, and ii. SEM image of the patterned structure iii. EDX element mapping showing the gold surface from the laser exposed region indicating removal of a-Sb$_2$S$_3$. iv. EDX element mapping confirming the presence of elemental sulphur in the a-Sb$_2$S$_3$ layer surrounding the print.

In 2017, Xiaomin Wang et al. demonstrated the thermal decomposition of Sb-BDCA for the fabrication of Sb$_2$S$_3$ thin films[21]. Subsequently, Wei Wang et al. extended the use of metal-BDCA complexes to direct lithographic patterning of metal chalcogenide nanostructures[12]. It was found that under fs pulsed irradiation at 780 nm, Sb$_2$S$_3$ can be dissolved in the presence of BDCA (Scheme. 1c). This enables a selective removal (wet-etching) of Sb$_2$S$_3$ realized by steering the femtosecond laser focus. The positive-tone patterning process of Sb$_2$S$_3$ is illustrated in Figure 1a. First, a diluted Sb-BDCA solution is spin-coated onto the target substrate (Step 1 in Figure 1a). Following thermal annealing (Step 2 in Figure 1a), a compact a-Sb$_2$S$_3$ thin film with ~66 nm thickness is obtained on the substrate as a result of the decomposition of Sb-BDCA (Scheme. 1b). During the etching process (step 3 in Figure 1a), the BDCA etchant is placed between the a-Sb$_2$S$_3$ thin film and the high numerical aperture (NA = 1.4) objective, i.e., used as both etchant and immersion medium. The 780 nm fs-laser beam is tightly focused on the a-Sb$_2$S$_3$ film (Figure 1b). Under laser irradiation, the a-Sb$_2$S$_3$ film is selectively etched by the BDCA (Scheme 1c). Pattern writing is achieved by steering the laser focus along a pre-programmed path, i.e., controlled by computer-guided galvo mirrors. Finally, in the development step (Step 4 in Figure 1a), the substrate is rinsed to remove residual BDCA etchant, completing the patterning process.

A well-defined pattern can be clearly seen from both optical (Figure 1c(i)) and SEM (Figure 1c(ii)) images. Energy-dispersive X-ray (EDX) spectroscopy was employed to analyze the elemental composition of the developed structure. The laser-exposed regions showed the presence of gold (Figure 1c(iii)), confirming the removal of a-$Sb_2S_3$, while the non-exposed areas contained sulfur (Figure 1c(iv)). Elemental mapping demonstrated that the "$Sb_2S_3$" pattern forms trenches, indicating the femtosecond laser assisted patterning method presented here as positive-tone lithography.

**2.2 Resolution and depth of femtosecond laser assisted wet-etching on a-$Sb_2S_3$**

The full width at half maximum (FWHM) of a tightly focused beam used for the inscription is determined to be ~ $\frac{\lambda_0}{2NA}$ (Abbe limit, $\lambda_0$ is the vacuum wavelength) in homogeneous medium[22]. However, in multilayer structures, the focused beam profile becomes complex due to standing wave patterns[23]. In practice, the laser intensity, the laser pulse property, scanning speeds, film thickness, material's etching threshold and film refractive index codetermine the laser writing resolution[24]. To obtain fine feature size, using high NA objective and decreasing film thickness are common methods to optimize spatial resolution. In our device, which comprises of a thin film (66 nm) of a-$Sb_2S_3$ layer sandwiched between $SiO_2$ and BDCA or Au and BDCA, the interaction between propagating plane waves at the $SiO_2$/a-$Sb_2S_3$ and Au/a-$Sb_2S_3$ boundaries, form a standing wave. This phenomenon is analytically described in ref. 25, where monochromatic waves interfering at substrate-thin film boundaries create spatially fixed-phase standing waves.

To study the shape formation mechanism and spatial resolution in femtosecond laser processing, we employed the Finite-Difference Time-Domain (FDTD) method for various substrates (more details given in the Part SI of Supplementary Information). Figure 2a shows the intensity distribution ($I = |\boldsymbol{E}|^2$) of a tightly focused (NA = 1.4), 780 nm linearly polarized Gaussian beam in the BDCA/a-$Sb_2S_3$/$SiO_2$ multi-layered structure. As illustrated in Figure 2a (top), the highest intensity is located 177 nm below the a-$Sb_2S_3$/BDCA interface. The intensity distribution along the optical axis (z direction) is also shown in Figure 2a (left). The source plane is scanned in the FDTD simulation setup to obtain the highest possible intensity ($I_{max}$) value and smallest possible FWHM value at a-$Sb_2S_3$/BDCA interface (at z = 0 nm position). We found through shifting the laser beam focus in the FDTD environment both in +z and -z direction, it is not possible to localise the strongest intensity inside the a-$Sb_2S_3$ layer (see Supplementary Figure S5). The positions of wave nodes are almost not changed by the source position (see Supplementary Movie S2), thus spatially fixed-phase standing waves were formed in the tightly focused scenario. This phenomenon can be attributed to the large refractive index mismatch of BDCA solution (n = 1.57) and a-$Sb_2S_3$ (n = 2.43), which leads to partial reflection (12.8 %) at the a-$Sb_2S_3$/BDCA interface. The superposition (interference) of incident (source) and reflected plane wave components leads to the maximum focal spot inside the BDCA layer (Supplementary Figure S5)[23]. The $I_{max}$ value in the a-$Sb_2S_3$/BDCA interface is 50.8 $V^2/m^2$ (Figure 2a (top)) with FWHM = 279 nm, which is close to the Abbe limit (278 nm)[26].

The situation is totally different in a BDCA/a-$Sb_2S_3$/Au multi-layered structure. As can be seen in Figure 2b, the strongest intensity is located at a-$Sb_2S_3$/BDCA interface. This can be attributed to the fact that the incident light gets mostly reflected (63.3%) at the 200 nm thick Au layer (see Supplementary Figure S2e and S6). A direct result of the strong reflection is an enhanced interference. The $I_{max}$ is recorded as 432.8 $V^2/m^2$ in the middle of a-$Sb_2S_3$ layer (Figure 2b(top)). In our FDTD implementation, the position of the maximum intensity while keeping the a-$Sb_2S_3$ film thickness at 66 nm, is always at a-$Sb_2S_3$/BDCA interface (see Supplementary Figure S6 and Movie S3), which shows another case of fixed-phase interference. The FWHM at the interface is recorded as 300 nm. Thus, based on the FDTD results, we can predict that the patterning on Au would require much less laser power than on $SiO_2$ due to the high laser power at a-$Sb_2S_3$/BDCA interface. However, the resolution should be better on $SiO_2$ substrate because of the relatively narrow beam width (FWHM).

Following the simulation results, we performed experiments by continuously exposing the sample with different laser powers, ranging from 4.4 to 17.6 mW under the same scanning speed (200 μm/s). A series of 26 1D-linear gratings were patterned on the $SiO_2$ and Au substrate (as indicated in the inset of Figure 2c and 2d with laser power increasing from bottom to top (vertically oriented gratings) and left to right (horizontally oriented gratings). Essentially, it can be pointed out from the optical image of the bottom to top (for vertically oriented gratings) and left to right (for horizontally oriented gratings). Essentially, it can be pointed out from the optical image of the patterned gratings that deeper grooves appear at higher laser power. The trench depth was measured with atomic force microscopy (AFM) as displayed on Figure 2c and 2d. In Figure 2e and 2f, we plot the measured line width and trench depth as a function of different laser powers. From these plots, we can see that the trench depth can be controlled via laser power, showing the possibility of direct write grayscale (2.5D) lithography [27]. Following the trend of the plot it can be observed that for lower laser powers, the ratio between the trench depth and the line width varies greatly. However, the ratio becomes constant at higher laser powers indicating a stable morphological control. On $SiO_2$, the required laser power for 100% penetration is 17.4 mW, compared with 8.6 mW on Au. This phenomenon can be explained by the numerical simulation data, where we predict that with the same $I_{max}$, the required power on Au is clearly lower than on $SiO_2$.

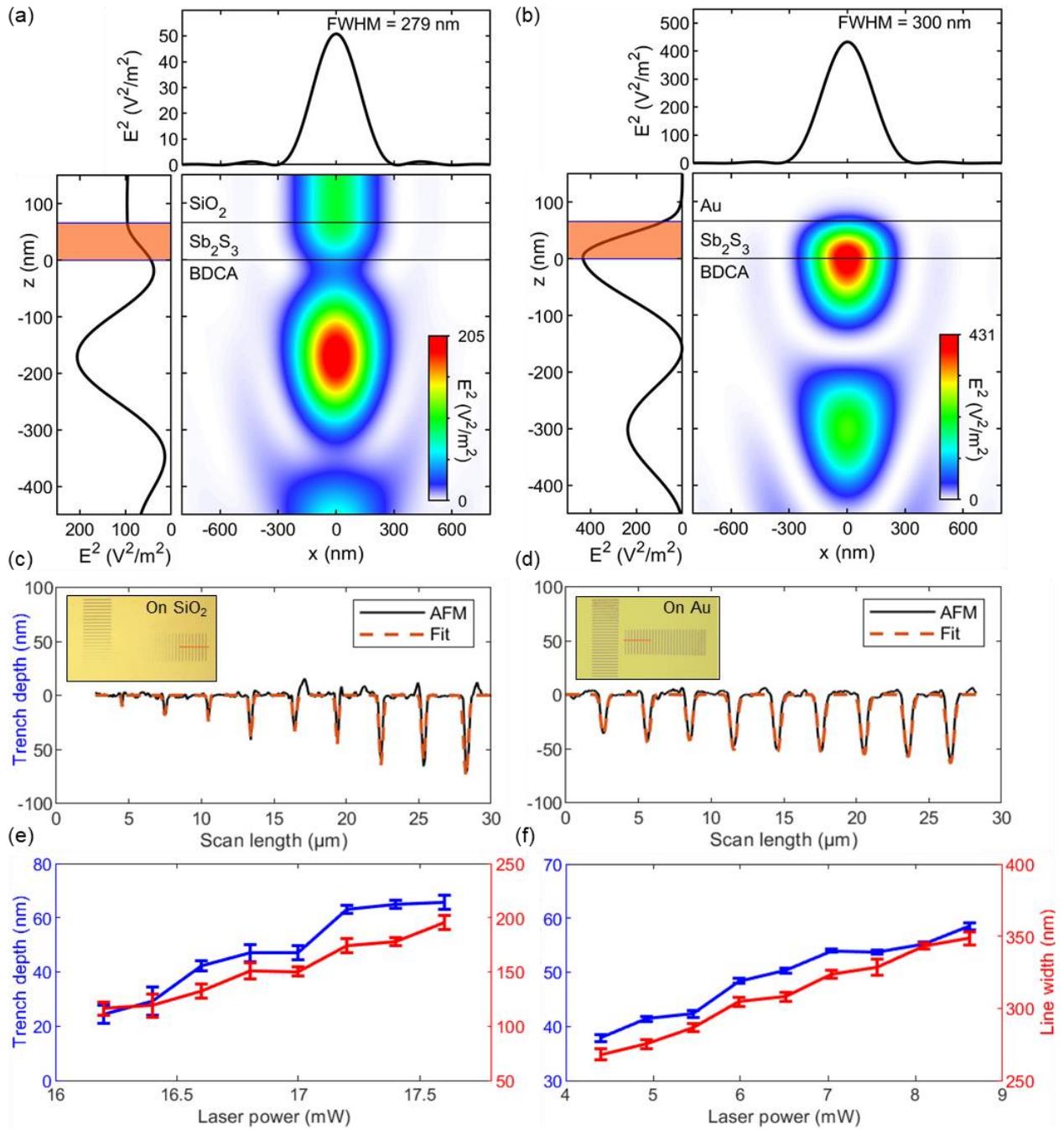

**Figure 2. FDTD simulations and AFM based depth characterization of linear grating structures on SiO$_2$ and Au substrate.** Intensity distribution along the XZ plane sectioning the BDCA/a-Sb$_2$S$_3$/SiO$_2$ (a) and BDCA/a-Sb$_2$S$_3$/Au (b) multi-layered substrate. The simulated maximum intensity ($I_{max}$) and FWHM at the BDCA/a-Sb$_2$S$_3$ interface (z = 0 nm) for the BDCA/a-Sb$_2$S$_3$/SiO$_2$ are 50.8 V$^2$/m$^2$ and 279 nm, and for BDCA/a-Sb$_2$S$_3$/Au are 432.8 V$^2$/m$^2$ and 300 nm. The trench depth vs scan length plot based on the collected AFM scan data on SiO$_2$ (c) and Au (d) substrate is shown. Optical BF image of a series of 26 linear gratings patterned both in vertical (left) and horizontal (right) direction on SiO$_2$ (inset, c) and Au (inset, d) substrate with increasing laser power (LP) and constant scan speed of 200 µm/s. The red line displayed in both inset indicates the AFM scan region. Trench depth (blue) and line width (red) of trenches vs the incident laser power (Error bars represent standard error of the mean (SEM) calculated from multiple repeated measurements at each laser power value) on SiO$_2$ (e) and Au (f) substrate.

It is important to note that, given the maximum trench depth, the line width achieved on the $SiO_2$ substrate (Figure 2c) was significantly smaller compared to the one on the Au substrate (Figure 2d). This indicates that structures with higher resolution can be attained with the a-$Sb_2S_3$ on $SiO_2$ substrate. This fact is vital for the application on fiber end faces [28,29]. According to Figure 2e, the smallest line width (100% penetration) achievable is 178 nm on $SiO_2$, this is about 64% of the Abbe limit. This result is not surprising, as it is very common to obtain sub-diffraction limited resolution with direct laser writing [9,23,30-32]. The laser intensity distribution, scanning speeds, film thickness and film refractive index are the major factors affecting the laser writing resolution[24]. The smallest width (100% penetration) is ~350nm on Au (Figure 2f). The resolution on quartz is better than on gold, which also agrees well with the FDTD simulation results.

**2.3. Demonstration with various test patterns:**

To demonstrate the fabrication potentials, special test patterns were patterned on both the transmissive ($SiO_2$) (Figure 3a) and reflective (Au) (Figure 3b) substrates. These nanopatterns were created with pre-defined parameters containing variable laser powers and scan speeds (details regarding the nanopatterns is provided in Supplementary Figure S8). Here we present single items from each individual structure for the ease of representation. The nanostructured patterns as portrayed in Figure 3a(i-iii) and Figure 3b(i-iii), were used from Nanoscribe's test pattern library. The grating structure shown in Figure 3a(i) and Figure 3b(i) contain variable periodicities used for measuring lateral resolution in 2D. The periodicity range spanned between 0.5, 1, 1.5 and 2 μm from bottom to top in the Y-direction and from left to right in the X-direction. The highest achievable resolution can be quantified by studying the smallest periodicity that can be separately resolved. It can be observed that the gratings patterned on $SiO_2$ substrate (Figure 3a(i)) have better resolution compared to the Au substrate (Figure 3b(i)). This can be explained once again from the resolution and depth characterization test given in Figure 2. The line patterns appear identical along X and Y-direction on both substrates. Figure 3a(ii) and Figure 3b(ii) demonstrates the trajectory test pattern which is used to test the accuracy with which the programmed coordinates are imaged into the substrate during the printing process. At high scan speeds during the printing process, a 90° corner is very difficult to shape. Hence this test pattern gives a good grasp on how structures with varying morphology would appear during/after printing. It can be observed from both the test patterns on $SiO_2$ and Au substrates that a good control over varying trajectories can be achieved.

Next, we fabricated structures resembling a bull's-eye grating or circular Bragg grating (Figure 3a(iii) and Figure 3b(iii)), comprising concentric rings surrounding a central aperture, on both $SiO_2$ and Au substrates. These structures were used to evaluate the patterning quality within the Galvo scan field, demonstrating stable performance across varying ring radii. Notably, the ability to create such geometries enables the fabrication of both transmission- and reflection-based flat optics, including metalenses, gratings, diffractive optical elements (DOEs), and zone plates—offering a promising, miniaturized alternative to bulky and expensive conventional refraction-based optics[16,33,34]. Such structures can also be employed for improved light coupling into optical fibres [28,29]. Furthermore, the utility of this technique can be extended to create complex photonic structures on non-wafer substrates such as optical fibers, which is difficult to realise conventionally due to complicated multi-step fabrication process[35]. To further demonstrate the potential of our positive-tone printing approach in metasurface fabrication with variable periodicities, we patterned azimuthally chirped gratings (ACGs) on both SiO$_2$ and Au substrate. ACGs disperse broadband white light illumination into its constituent wavelengths via its azimuthally varying periodicity (as seen in Figure 3a(iv,v) and Figure 3b(iv,v)), effectively displaying its colour sorting quality.

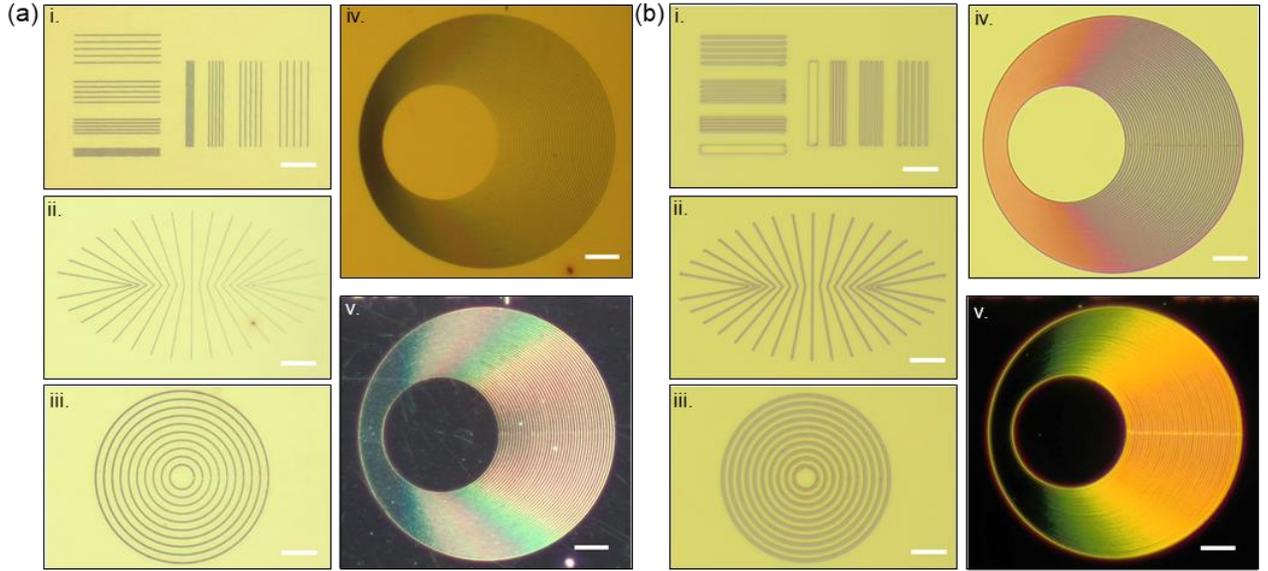

**Figure 3. Test pattern for evaluation of various geometric patterns on (a) transmissive (SiO$_2$) and (b) reflective (Au) substrates.** i. 2D gratings with changing periodicities patterned in both horizontal and vertical direction ii. Trajectory test structure with varying corner angles iii. Concentric rings showing bull's eye type grating structures iv. Bright field optical image of ACG structure with azimuthal angle dependant varying periodicity, ranging from 1000nm at the 0° azimuthal angle till 200 nm at the 180° azimuthal angle v. Dark field optical image of the ACG displaying dispersion spectrum. Scale bar = 10μm for all images.

## 2.4. Demonstration with a-Sb$_2$S$_3$ Fresnel Zone Plates

Refractive lenses are widely used in optical systems and daily life, but their bulky size renders them incapable of meeting the growing demand for miniaturization in integrated optical systems [36-38]. The Fresnel Zone Plate (FZP) is a typical planar diffractive optical device that consists of alternate concentric rings of optical transmittances (amplitude type) or various thicknesses (phase type) [39]. The light beams transmitted by FZPs interfere constructively at the designed points (see Figure 4a). Various DOEs and hybrid FZPs have been recently fabricated through holographic femtosecond lasers on metal films [40], direct laser write assisted etching of graphene oxide [41] and EBL assisted monolayer MoS$_2$ based reflective metalens [42]. However, the aforementioned methods often involve fabrication techniques that have slow throughput, limited resolution and low surface quality [13]. Our approach of wet-etch femtosecond laser assisted fabrication of Sb$_2$S$_3$ mitigate the fabrication bottlenecks by providing a method to pattern mask-free, high resolution and single layer FZPs.

For the design of a binary FZP, the radial spatial patch path difference equation is given by [39]:

$$\sqrt{f^2 + r_n^2} - f = n\lambda \quad (1)$$

where, $f$ is the designed focal length, $n$ is the number of n$^{th}$ zone (n = 1, 2, 3, 4….), $r_n$ is the radius of the n$^{th}$ zone, $\lambda$ is the wavelength in the designed working medium. With a fixed $f$ value, $n$ defines the resolution limitation (numerical aperture) of the FZP. The binary zone plate in our case represents the most basic two-level approximations. As depicted in Figure 4a, each zone has an additional $2\pi$ phase shift. Solving Equation (1), we can obtain:

$$r_n = \sqrt{\frac{n^2\lambda^2}{4} + nf\lambda} \tag{2}$$

Based on Equation (2), we created three a-$Sb_2S_3$ FZPs with $\lambda = 633$ nm, $f = 800, 600, 400$ μm and $n = 18, 20$ and $25$ respectively. The optical setup used for imaging using the FZPs is shown in Figure 4b. The setup used for optical characterization consisted of an LED illumination source, followed by the object to be imaged (the target). To showcase the potential of the concept, we used a ruler with the word "**IPHT**" written on top (Figure 4c). The $SiO_2$ substrate containing the a-$Sb_2S_3$ FZPs were placed at 11.5 cm distance away from the target. A 50X, 0.7 N.A. objective lens was placed after the image plane followed by a tube lens through which the rays were focussed on a CCD camera for further characterization of the image. As seen on Figure 4d, the target was successfully imaged by the FZPs.

The imaging with a lens, it is very common to use the lens equation given by:

$$\frac{1}{f} = \frac{1}{u} + \frac{1}{v} \tag{3}$$

where $u$ and $v$ are the target and image. The magnification $M = \frac{v}{u}$ defines ratio of the height of an image to the height of an object. Based on Equation (3), we can thus calculate the ratio of the image height as 2:1.5:1 for the $f = 800, 600, 400$ μm FZPs, respectively, which is in good agreement with the experiment results shown in Figure 4d.

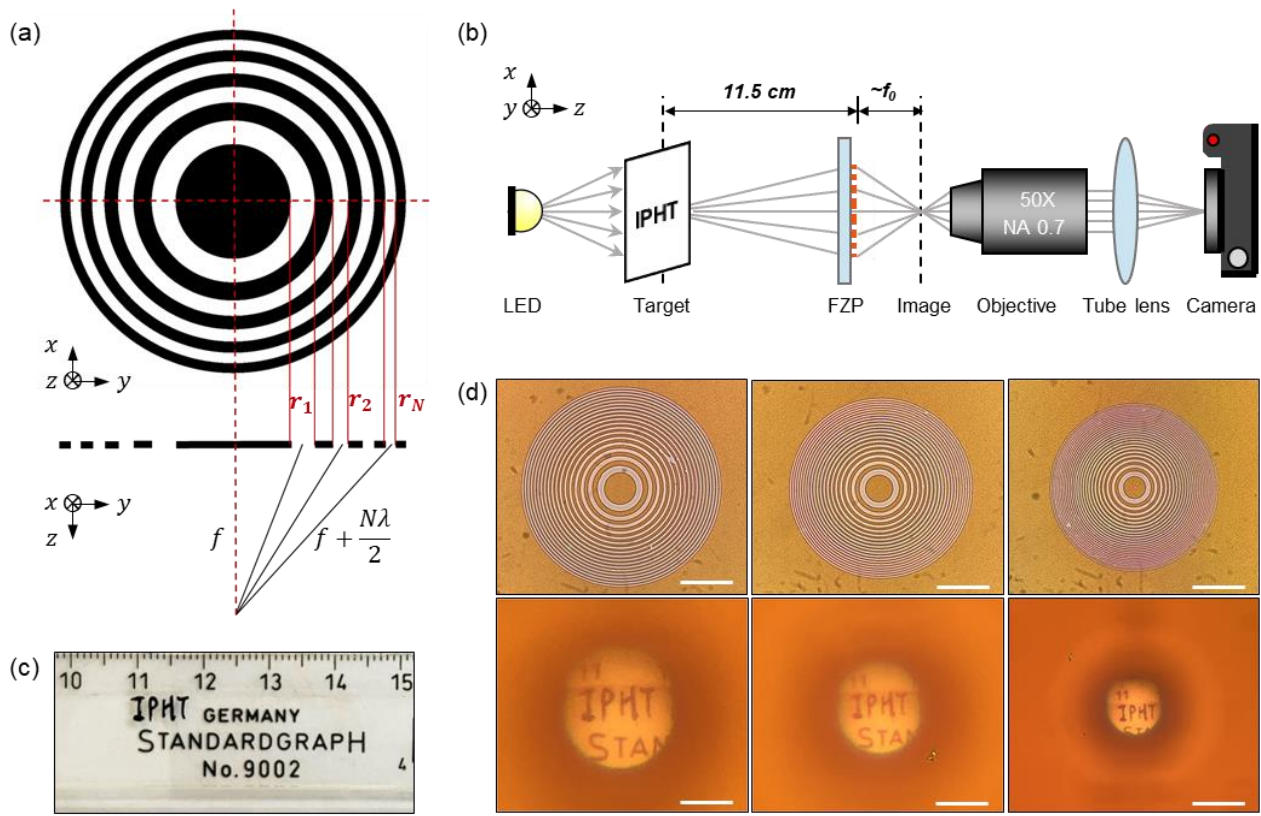

**Figure 4. Fresnel Zone Plates (FZP) designed for various focal lengths operating at the wavelength of 633nm.** (a) Design principles behind constructing an FZP. (b) Schematic illustration of the optical setup used for imaging the target. (c) A centimeter ruler with the handwritten text "IPHT" written on top is used as the target object. (d) Top: various FZPs with focal length f = 800μm, 600μm and 400μm were fabricated. Bottom: the images obtained with each lens is shown simultaneously. Scale bar: 50 μm.

## 3. Conclusion

In this work, positive-tone femtosecond laser assisted wet-etch lithography of a-$Sb_2S_3$ is realized for the first time. The flexibility of this fabrication method enables $Sb_2S_3$ based micro- and nano-optical device fabrication on both transmissive and reflective substrates. We showed that femtosecond laser induced $Sb_2S_3$ wet-etching in the BDCA is the decisive mechanism for the direct patterning process. Our results clearly demonstrate the capabilities of the femtosecond laser assisted wet-etching method by printing linewidth down to 178 nm while maintaining controlled depth. FDTD was used for the simulation of the focused light intensity distribution during the pattering process on quartz (transmissive) and gold (reflective) substrates. The simulation results predict that as a result of more efficient energy concentration, the pattering on gold substrate would require much less power than on the quartz substrate. However, the printing resolution on quartz substrate should be better than on gold given the narrow beam width. The experimental observation agrees well with the simulation results. Various test patterns, especially azimuthally chirped gratings were fabricated as demonstration. In the end, we fabricated a-$Sb_2S_3$ FZPs with varied focal length, demonstrating imaging capabilities. We believe that the reported method enables a rapid and flexible design and fabrication of lab-on-chip optoelectronic devices in a straightforward manner. This technology also facilitates the fabrication of nanostructures on unconventional substrates, such as optical fiber end faces or directly on camera sensors. This unique capability expands the range of potential applications, including nanostructure-enhanced light coupling into fiber cores[28] and lensless microscopy[43].

## 4. Materials and methods:

*Materials:* The $Sb_2S_3$ thin film was synthesized by well-established protocols from previous publications [21]. Antimony (III) oxide ($Sb_2O_3$, 99.99%) was obtained from thermo scientific, 1-butylamine (($CH_3(CH_2)_3NH_2$), 99.5%) and carbon disulphide ($CS_2$, 99.9%) was purchased from Sigma-Aldrich, and ethanol ($CH_3CH_2OH$, 99.99%) was acquired from Merch KGaA. For the gold substrates, 200nm of gold was sputtered onto 1mm thick fused silica ($SiO_2$) substrate with lateral dimension of 18 x 18 $mm^2$.

*Preparation of Sb-BDCA precursor solution:* The BDCA etchant is formed by mixing 2 mL 1-butylamine and 1.5 mL carbon disulphide in 2 mL ethanol. For synthesizing the Sb-BDCA acid, carbon disulphide (1.5 mL), Antimony (III) oxide (1.0 mmol) and ethanol (2.0 mL) were mixed in a 100 mL three-neck vial and were stirred continuously at room temperature assisted by magnetic stirring. Into this solution, 1-butylamine (2mL) was gradually added into the vial, following which the solution was stirred continuously for a duration of 12 hours until a uniform and clear solution was formed. In order to synthesize $Sb_2S_3$ thin film of desired thickness, the Sb-BDCA complex solution was combined volumetrically by mixing ethanol (2 mL) to Sb-BDCA complex (1 mL). It is worth mentioning here that different thicknesses of $Sb_2S_3$ thin films can be manufactured using this method by controlling the volumetric ratio between Sb-BDCA and ethanol. Afterwards, the precursor solution was spin coated (Laurell spin coater, model WS-650HZB-23NPPB) on top of fused silica substrate and gold-coated fused silica substrate separately at 8000 rpm for 30 seconds. The high rotation speed ensures uniform and homogeneous coating of the Sb-BDCA film on the substrates [11]. After spin coating, the substrates were annealed on a hot plate at 180°C for 10 minutes, following which both substrates deposited with amorphous $Sb_2S_3$ thin films were obtained.

*Wet-etch femtosecond laser assisted lithography:* All structures were programmed using Describe programming language (proprietary software from Nanoscribe GmbH). All patterns were patterned using Nanoscribe's Photonic Professional GT2 (PPGT2) at room temperature with a pulsed femtosecond erbium-doped fibre laser (Toptica Photonics AG, Munich, Germany) with centre wavelength at 780 nm. The pulse duration of the laser is 80-100 fs and the repetition rate 80 MHz. A 63× Oil DIC Plan-Apochromat objective lens with Numerical Aperture (NA) 1.4 from Carl Zeiss was immersed directly into BDCA acid on top of substrate containing a-$Sb_2S_3$ film. We have to emphasize that the index matching for the immersion lens is additionally satisfied as the refractive index of the BDCA solution is measured as n=1.5749 at 588 nm vacuum wavelength, which is close to the value of designated immersion oil (n=1.518). After printing, the substrate with the structure was taken out for development process. The residual BDCA etchant was washed away by isopropanol followed by sonication in an isopropanol bath for 30 seconds. Afterwards, the substrate was quickly dipped into Novec™ solution (from 3M™) and air dried. For the FZP fabrication on quartz, the scan speed is set at 75 µm/s and laser power is 15.4 mW.

*Device and material characterization:* The SEM measurements were performed with a scanning electron microscope Tescan LYRA (Tescan, Brno, Czech Republic). The energy of the exciting electrons was set to 5 keV in most cases. All energy dispersive X-ray (EDX) analyses were done using a state of the art 30 mm$^2$ silicon drift detector (SDD) by BRUKER (BRUKER Nano GmbH, Berlin, Germany) and the Esprit spectra evaluation software package. The specified energy resolution of the detector at 5.9 keV (Mn-Kα) is 129 eV. The AFM measurements were performed with a BRUKER Dimension Edge system (BRUKER AXS, Karlsruhe, Germany) operated in contact mode. Supersharp SNL (silicon tips, Bruker) probes with a specified tip radius of 2 nm were used.

*Numerical simulations:* In order to characterize the electric field intensity distribution for a-$Sb_2S_3$ film deposited both on $SiO_2$ and Au substrates, finite-difference time-domain (FDTD) simulation (Ansys Lumerical Inc.) was setup. Simulation setups consisting of a top layer of BDCA with an index value n = 1.5749, followed by a layer of a-$Sb_2S_3$ with thickness 66 nm on a $SiO_2$ substrate and Au substrate was created separately in a FDTD environment. The source was set as a X-polarised converging Gaussian beam source centred at a wavelength of 780nm with N.A = 1.4 and injected along the +Z direction. A boundary condition of "anti-symmetric" is set to x-min and "symmetric" is set to y-min, which saves 75% of the required simulation time and memory. The power of the source is measured as $2.37 \times 10^{-15}\ watt$ based on the far-field integration (using "farfield3dintegrate" function from Lumerical). This source was used for all the FDTD simulations presented in this work. In order to view the electric field intensity distribution, a 2D frequency and power monitor was set cross-sectioning the entire device along the XZ plane. A perfectly matched layer (PML) were used in the Z direction to absorb the outgoing waves.

## 5. Acknowledgements:

VD, JSH and AD thank for financial support from The DFG via SFB 1375 NOA (Project No.: 398816777). JSH and AD appreciates support via the IRTG 2675 Meta-Active (Project No.: 437527638). WW thanks for financial support through the IPHT Innovation Project 3D-HiRes 2021/2022 (Project No.: K690082) and the DFG via CRC-TRR 234 CataLight (Project No. 364549901). AD and WW thank Dr. Oleg A. Egorov and Dr. Jinxin Zhan for their help with the simulation. AD and WW thank Prof. Dr. Lukas Novotny for the discussion. All the simulations were performed on the HPC-cluster supported by EFRE Programm and Freistaat Thüringen, projects EU-0V/2020-59 (2019 FGI0017) and EU-0V/2023-1 (2022 FGI 0004). This work is further supported by the BMBF, funding program Photonics Research Germany ("LPI-BT1-FSU", FKZ 13N15466) and integrated into the Leibniz Center for Photonics in Infection Research (LPI). The LPI initiated by Leibniz-IPHT, Leibniz-HKI, UKJ and FSU Jena is part of the BMBF national roadmap for research infrastructures.

## 6. Author contributions:

AD and WW conceived the idea. AD and WW synthesized the material and designed the experiment. AD and WW performed the direct laser writing process and optical inspection. AD, XW and WW performed the FDTD simulation. HS, AB and UH prepared the substrates. Andrea Dellith and JD performed the AFM, SEM and EDX characterization. TW measured the refractive index of a-$Sb_2S_3$. MZ and MAS were involved in the design of the FZPs. AD and WW wrote the manuscript and all authors contributed to the writing of the manuscript. JSH and VD supervised the study.

## 7. Conflict of Interest:

The authors declare no competing interests.

**8. Data Availability Statement:** The data that support the findings of this study are available from the corresponding author upon reasonable request.

# Supplementary Information

**Positive-tone Nanolithography of Antimony Trisulfide with Femtosecond Laser Wet-etching**

**Part SI: Tightly focused beam in the presence of multilayered structure**

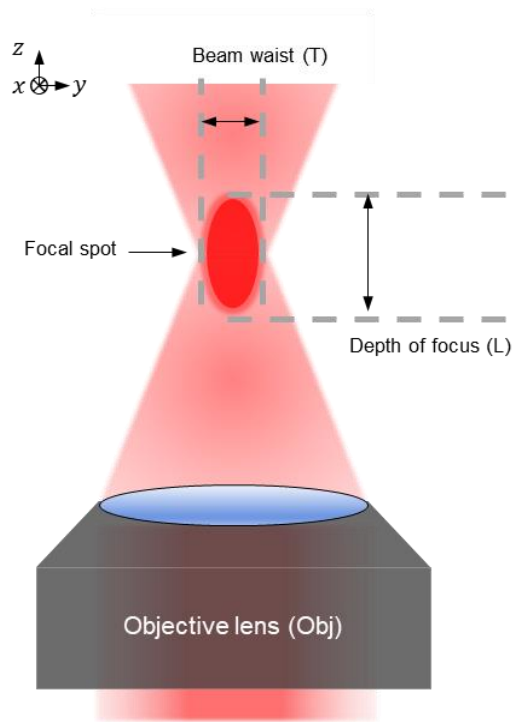

**Figure S1.** Light distribution in the focal region of a high NA OL. The focusing spot sizes, T and L, are defined at full width at half maximum (FWHM) of its peak intensity along the transverse (beam waist) and longitudinal (depth of focus) directions, respectively.

Since its invention in the early 17th century, the optical microscope [44] has attracted significant attention from researchers due to its ease of operation, precise focusing capabilities, and broad range of applications. Optical microscopy relies on a system of combined optical lenses to investigate materials of interest. The recent advent of high numerical aperture (NA) objective lenses (OLs) has enabled light to be focused into sub-micrometer [44] or even nanoscale spots [45], expanding its utility across diverse fields, including nanofabrication [46,47], high-resolution imaging [48], optical data storage [49], optical acceleration, and optical trapping [50]. Conventional optical microscopy primarily operates on a one-photon absorption (OPA) mechanism. However, under tight focusing conditions, the resulting highly confined focal spot is highly sensitive to various parameters, including the characteristics of the incident light, the properties of the objective lens, and the optical parameters of the surrounding medium. A deeper understanding of these factors, along with the ability to tailor focal spot characteristics, is crucial for optimizing applications. In this supplementary information, we explore the behaviour of tightly focused laser spots in multi-layered structures. While our discussion focuses on the positive printing of a-$Sb_2S_3$ on different substrates, we anticipate that the insights gained can be broadly extended to other applications.

Diffraction, a fundamental manifestation of the wave nature of light, governs the behaviour of tightly focused beams. Figure S1 illustrates the propagation of a light beam focused by an aberration-free, high-numerical-aperture (NA) objective lens (OL). In the focal region, rather than converging to a singular point, the focal spot assumes an ellipsoidal shape with rotational symmetry about the optical axis. The dimensions of this focal spot are characterized by the transverse (T) and longitudinal (L) extents, with their ratio (L/T) often defined as the aspect ratio (AR) of the laser spot. The finite size of the focal spot arises from diffraction effects as light propagates through the lens aperture. The lens pupil acts as a diffracting aperture composed of an infinite number of diffracting points. Near the focal region, the electromagnetic field (EM) distribution results from the superposition of all diffracted rays emerging from the aperture.

The size and shape of the focal spot have direct implications for the resolving power of an optical microscope. Imaging a point object can be modelled as a Dirac delta function (often described as $\delta$-function)—yields a diffraction-limited focal spot described by the point spread function (PSF). When two point objects are positioned within the focal region at sufficiently small separations, their PSFs overlap, preventing their distinction as separate entities. Similarly, in optical lithography—such as in this work—the minimum feature size that can be fabricated is constrained by the transverse and longitudinal dimensions (T, L) of the focal spot. Mathematically, optical resolution is often described by the criteria established by Abbe [22] and Rayleigh [51]. While both formulations are conceptually similar, they differ in the numerical coefficients due to distinct definitions of the minimum resolvable distance between two closely spaced objects. According to Rayleigh's criterion, the transverse and longitudinal resolution limits are given by [51]:

$$T = \frac{0.61 \lambda_0}{NA} \quad \text{S(1)}$$

$$L = \frac{2 \lambda_0}{NA^2} \quad \text{S(2)}$$

where $\lambda_0$ is the wavelength of light in vacuum, NA is the numerical aperture of OL, which is defined as:

$$NA = n \sin \theta \quad \text{S(3)}$$

where $n$ is the refractive index of the medium in which the light propagates, and $\theta$ is half of the maximum angle through which the light is focused into the focal region (Figure S2a). According to these formulae, the resolution of an optical imaging system is referred to light wavelength $\lambda$, and the NA of the used OL. Smaller focusing spot and better resolution can be obtained with higher NA OLs. To be a high NA, one can increase the $\theta$-angle or use an immersed medium with higher refractive index, $n$. The most commonly used immersion medium is a transparent oil with $n = 1.518$ (at 588 nm vacuum wavelength), such as Immersol 518 F from Carl Zeiss GmbH. In this work, the refractive index of BDCA acid is 1.5749 (measured with Abbe refractometer [26] from VEB Carl Zeiss Jena, see Table S1).

In practice, one critical problem concerning optical microscopy application is aberration caused by the refractive index mismatch media, which is primary composed of spherical aberration [52]. As illustrated in Figure S2a, when light is tightly focused through refractive mismatch medium, it does not focus at the designed focal point $f_0$ (assume the OL is design at $n_1$ media). Depending on the refractive index of the first and second media (with refractive index $n_1$ and $n_2$, respectively), the focal spot shifts either at the down ($F_1$) or up ($F_2$) of its original focal point $f_0$. According to the work of Joel et al. [53], the refractive index of plane-parallel sample can be determined based on such focal point displacement. Such focal point displacement can be easily explained by Snell's law and ray optics. The Snell's law [54] states that, for a given pair of media, the ratio of the angle of incidence ($\theta_1$) and angle of refraction ($\theta_2$) is equal to the ratio of the refractive indices of the two media:

$$\frac{\sin \theta_1}{\sin \theta_2} = \frac{n_2}{n_1} \quad \text{S(4)}$$

For the case of focusing at BDCA/a-Sb$_2$S$_3$ interface (see Figure S2c), as $n_{BDCA} < n_{a-Sb2S3}$, according to Eq. S(4), the angle of refraction should be smaller than the angle of incidence ($\theta_2 < \theta_1$), as depicted in Figure S2b. Thus, one can expect up shifted and elongated focal spot in this case. Besides, due to the presence of the interface, an undesired phase aberration is introduced at the original focal plane [55]:

$$\Phi(\theta_1, \theta_2, f_0) = -f_0(n_1 \cos \theta_1 - n_2 \cos \theta_2) \quad \text{S(5)}$$

$$t_s = \frac{2 \sin \theta_2 \cos \theta_1}{\sin(\theta_1 + \theta_2)} \quad \text{S(6)a}$$

| Refractive index (@780 nm) | n | k |
|---|---|---|
| BDCA | 1.575 | 0.000 |
| a-$Sb_2S_3$ | 2.431 | 0.019 |
| Au | 0.147 | 4.741 |
| $SiO_2$ | 1.454 | 0.000 |

**Table S1.** Refractive index (n+ik) of the materials at 780 nm vacuum wavelength for the FDTD simulation.

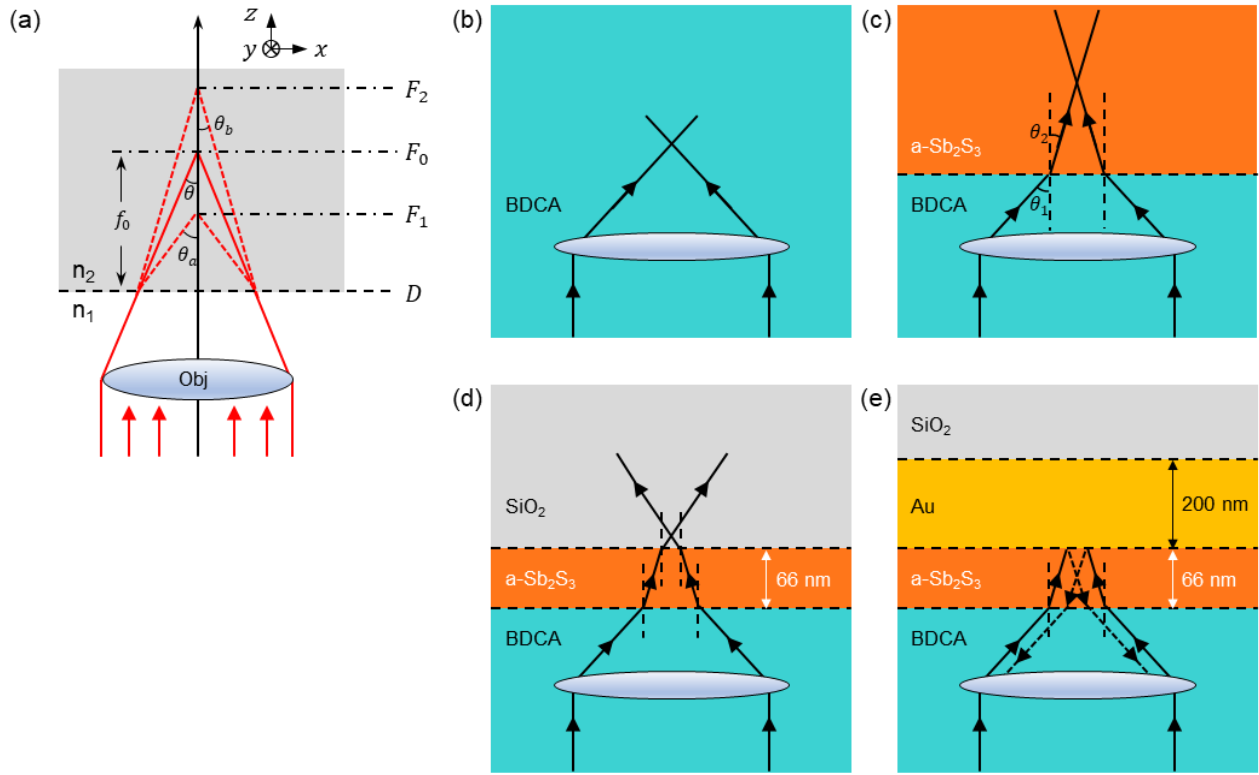

**Figure S2.** (a) Schematic illustration of the propagation of a tightly focused beam in the refractive index mismatch medium. $n_1$, $n_2$ are the refractive indices of the first and second media, respectively. $f_0$ is the focal point in the case of homogeneous medium (when $n_1=n_2$), and $F_1$ and $F_2$ the focal points when $n_1 < n_2$ and $n_1 > n_2$, respectively. $D$ is the interface between two media. (b) Focused beam in homogeneous BDCA medium with no refractive index mismatch. (c) Focused beam at BDCA/a-$Sb_2S_3$ interface, where $n_{BDCA} < n_{a-Sb2S3}$ with the actual focal point shifted up. (d) Focused beam at BDCA/a-$Sb_2S_3$/$SiO_2$ interface, with the actual focal point shifted up. (e) Focused beam at BDCA/a-$Sb_2S_3$/Au (200 nm) interface, where the actual focal point is reflected by the Au layer. For (c) and (d), the thickness of a-$Sb_2S_3$ and Au are 66 nm and 200 nm, respectively For the case of focusing at BDCA/a-$Sb_2S_3$/$SiO_2$ interface (see Figure S2d), the high index a-$Sb_2S_3$ is sandwiched between two low index media (BDCA and $SiO_2$). According to Snell's law, the focal point is expected to shift upward qualitatively.

$$t_p = \frac{2\sin\theta_2 \cos\theta_1}{\sin(\theta_1 + \theta_2)\cos(\theta_1 - \theta_2)} \qquad \text{S(6)b}$$

$$r_s = \frac{n_1 \cos\theta_1 - n_2 \cos\theta_2}{n_1 \cos\theta_1 + n_2 \cos\theta_2} \qquad \text{S(6)c}$$

$$r_p = \frac{n_1 \cos\theta_2 - n_2 \cos\theta_1}{n_1 \cos\theta_2 + n_2 \cos\theta_1} \qquad \text{S(6)d}$$

In 1950's, Richard and Wolf [56] formulated a comprehensive mathematical representation of the electromagnetic (EM) field distribution in the focal region of a high numerical aperture (NA) objective lens. Their model accounts for the vectorial nature of the EM field and is derived using the vectorial Debye approximation. Following Richard and Wolf, Novotny and Hecht [57] extended the approach to derive analytical expressions for strongly focused EM fields near dielectric interfaces. Their method incorporates Fresnel transmission and reflection coefficients (Eq. S(6)) within the vectorial Debye approximation. The Novotny and Hecht method also enables the representation of evanescent (near-field) components at the glass/air interface due to total internal reflection. In this work, instead of adopting the Novotny and Hecht method, Finite Difference Time Domain (FDTD) [58] is used for the simulation of the tightly focused field in various multi-layered devices. For the case of focusing at BDCA/a-$Sb_2S_3$/Au/$SiO_2$ interface, (see Figure S2e), the situation is the most complicated. Given the substantial thickness of the Au layer (200 nm), light transmission is negligible due to gold's high extinction coefficient (see Table S1). Instead, the majority of incident light is expected to be reflected, as thick plasmonic materials such as Au, Ag, and Al effectively impose Dirichlet boundary conditions [58]. FDTD is a viable approach for analyzing the electromagnetic field distribution in this case.

The single wavelength (frequency) FDTD simulations were carried out using Lumerical FDTD Solvers. Figure S3-S6 show the FDTD simulation results. The refractive index (see Table. S1) of the materials were imported into Lumerical directly as the device's optical data. An incident Gaussian wave featured as 2 μm ($f_0$) distance below the focal plane (NA=1.4, type "thin lens") along the forward z-axis direction (see Figure S3a) were used as source. The source is also linearly polarized (along x-axis). A boundary condition of "anti-symmetric" is set to x-min and "symmetric" is set to y-min, which saves 75% of the required simulation time and memory. The power of the source is measured as $2.37 \times 10^{-15}$ $watt$ based on the far-field integration (using "farfield3dintegrate" function from Lumerical). This source was used for all the FDTD simulations presented in this work. As depicted in Figure S3a, two 2D frequency domain monitors were place at the front and end of the z-axis to record the power transmission (noted as Tran) and reflection (noted as Ref). The transmission and reflection powers are calculated based on the integration of the Poynting vector. A ".lsf" script was written to generate different source z-position offset (noted as $d$) using the same source automatically. Another ".lsf" script was written to export desired data into the ".mat" MATLAB format automatically. All the data analysis and plots were performed based on MATLAB (version R2024a).

To assess the quality of the source, we conducted an FDTD simulation of tightly focused light in homogeneous BDCA media. As shown in Figure S3b, the focal spot in homogeneous media exhibits near-perfect symmetry across the xz-plane (with z = 0 nm), indicating negligible spherical aberration [59,60]. Moreover, nearly all incident light is transmitted (with an efficiency of 99.5%), and no reflection is observed, confirming that the source defined by Lumerical functions as a Total-Field /Scattered Field (TF/SF) source [58]. The TF/SF source technique injects a "one-way" field, eliminating backward-propagating waves and ensuring that 100% of the injected power is incident on the simulated device. Based on Figure S3c, the beam waist (T) is 300.6 nm. This value is slightly worse than the Abbe diffraction limit [22] with $0.5\lambda_0/NA \sim 278.6$ nm, but better than the Rayleigh's criterion (Eq. S1) [51] with $0.61\lambda_0/NA \sim 339.9$ nm. It is important to emphasize that both theoretical limits provide reasonable approximations of the focal spot size. However, for high-resolution imaging and nanofabrication applications, more rigorous methods—such as vectorial Debye theory (e.g., Richard-Wolf or Novotny-Hecht approaches) or finite element analysis—should be employed for precise calculations. Based on Figure S3d, the depth of focus (L) is 780.3 nm. This value is slightly smaller than the Rayleigh's criterion (Eq. S2) [51] with 795.9 nm. The aspect ratio (L/T) of the focusing spot is 2.6, which is in good agreement with previous report [52]. In summary, the Lumerical source exhibits exceptional quality due to its aberration-free nature, TF/SF implementation, and diffraction-limited PSF.

Figure S4 shows the FDTD simulation results of focusing at BDCA/a-Sb$_2$S$_3$ interface. As the working mechanism of the patterning is laser induced a-Sb$_2$S$_3$ etching with BDCA acid, we use the FWHM and intensity at BDCA/a-Sb$_2$S$_3$ interface ($z = 0$ nm) as the evaluation criteria. As depicted in Figure S4b, the beam waist does not exhibit drastic changes with source offset (noted as $d$) ranging from -475 nm to 525 nm (we set the default source position at $z = -1925$ nm). The intensity, on the other side, exhibits obvious changes with offset position. The intensity at BDCA/a-Sb$_2$S$_3$ interface with $d = 0$ nm is 71.8 $V^2/m^2$ (Figure S4e). In comparison, the maximum laser intensity is 124 $V^2/m^2$ (see Figure S3b) without interface. However, the maximum intensity in the device is 151.8 $V^2/m^2$, when positioned at $z = -150$ nm (in the BDCA layer), see Figure S7a. This phenomenon is a direct consequence of the superposition (interference) of incident (source) and reflected plane wave components. It is clear that the intensity distribution in the BDCA becomes periodic because of the interference. According to the frequency domain monitor, 3.7% of light power is reflected at the interface. In our FDTD implementation (see Movie_S1.avi), as a result of the interference, it is not possible to position the maximum intensity into the a-Sb$_2$S$_3$ layer. Thus, spatially fixed-phase standing waves were formed in the tightly focused scenario. It is interesting to notice that the elongated PSF along z-axis with large source offset value (Figure S4h), this is in good agreement of the first-order estimation of the focusing behavior shown in Figure S2c.

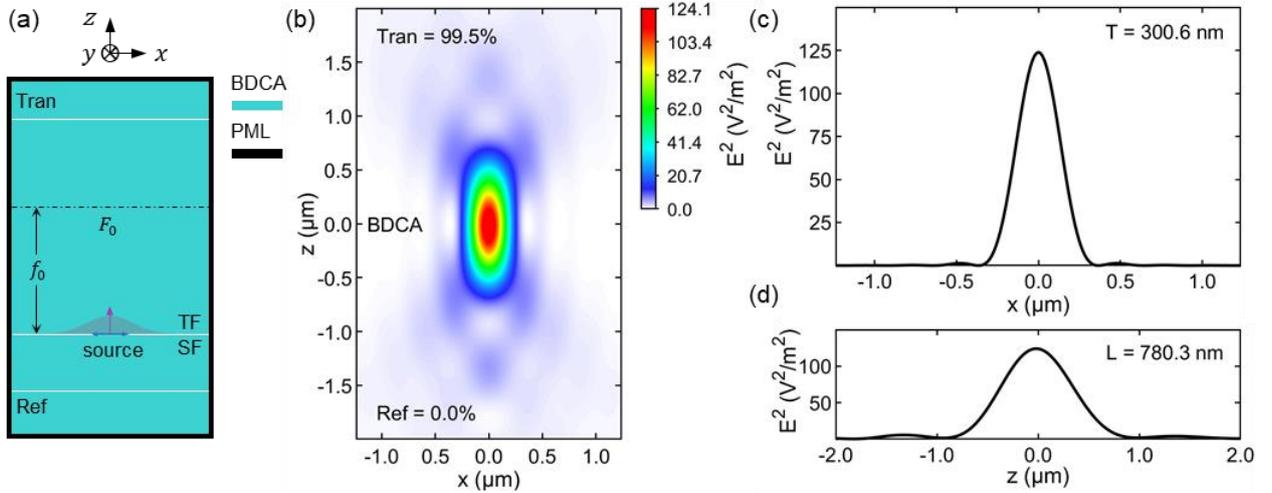

**Figure S3.** (a) xz-plane perspective view of the FDTD simulation setup of tightly focused beam in homogeneous BDCA media. $f_0 = 2\ \mu m$ is the distance between the focal plane ($F_0$) and the TF/SF source plane. Tran: Transmission monitor. Ref: Reflection monitor. PML: Perfect Matching Layer. (b) Longitudinal (xz-plane) intensity ($E^2 = E_x^2 + E_y^2 + E_z^2$) map of the focusing region. (c) The transverse intensity ($E^2$) profile. The beam waist (noted as T) is the FWHM of this profile. (d) The longitudinal intensity ($E^2$) profile. The depth of focus (noted as L) is the FWHM of this profile.

Figure S5 shows the FDTD simulation results of focusing at BDCA/a-Sb$_2$S$_3$(66 nm)/SiO$_2$ interface. As depicted in Figure S5b, the beam waist also does not exhibit drastic changes with source offset ranging from -475 nm to 525 nm. The intensity at $z = 0$ nm with $d = 0$ nm is 50.8 $V^2/m^2$ (Figure S5e). The reflection efficiency is 12.8%, which leads to the enhancement of interference compared with BDCA/a-Sb$_2$S$_3$ case. The maximum intensity in the device is 205.6 $V^2/m^2$, which still is positioned in the BDCA layer (at $z = -177\ nm$), see Figure S7b. The shift of the maximum intensity can be attributed to the interference of incident and reflected plane wave components from the BDCA/a-Sb$_2$S$_3$ and a-Sb$_2$S$_3$/SiO$_2$ interface. One can notice the wave front distortion caused by the multi-layers interface (Figure S5c). In our FDTD implementation (see Movie_S2.avi), it is also not possible to position the maximum intensity into the a-Sb$_2$S$_3$ layer as a result of the interference. Despite the low laser intensity at the interface, the FWHM of the beam is 279.2 nm (see top image of Figure 2a), which is the best among all the cases. Thus, one can predict the feature size of the printing on SiO$_2$ substrate should be the finest. In the future study, it is interesting to try

FDTD simulations with varied thicknesses of a-$Sb_2S_3$, as an effort to check whether is possible to position the maximum intensity at the BDCA/a-$Sb_2S_3$ interface.

Figure S6 shows the FDTD simulation results of focusing at BDCA/a-$Sb_2S_3$(66 nm)/Au(200 nm)/$SiO_2$ interface. No transmission can be detected by the frequency domain monitor, while most of the power is reflected. This result is in good agreement with the qualitatively estimation given in Figure S2e. A direct result of such strong reflection is the maximum 432.8 $V^2/m^2$ light intensity, which is the strongest for all the cases. What is more, the position of the maximum position is located at the a-$Sb_2S_3$/Au interface ($z = 0$ nm) position (see Figure S6e and Figure S7c). In our FDTD implementation (see Movie_S3.avi), the position the maximum intensity is always at $z = 0$ nm in the range of -275 nm to 325 nm source position offset. According to this simulation result, we found an optimized laser nanofabrication condition and it is the reason we use 66 nm thickness a-$Sb_2S_3$ in practice. The experimental result is in good agreement with FDTD simulation prediction, where the laser power for full penetration of the a-$Sb_2S_3$ layer is 8.6 mW on Au, which is only 48.9% of the power on $SiO_2$. Despite the high intensity at the BDCA/a-$Sb_2S_3$ interface, the FWHM of the field is 300.4 nm. Once can predict that the feature size on Au substrate should be larger than the one on the $SiO_2$ substrate, as a result of the increased FWHM value.

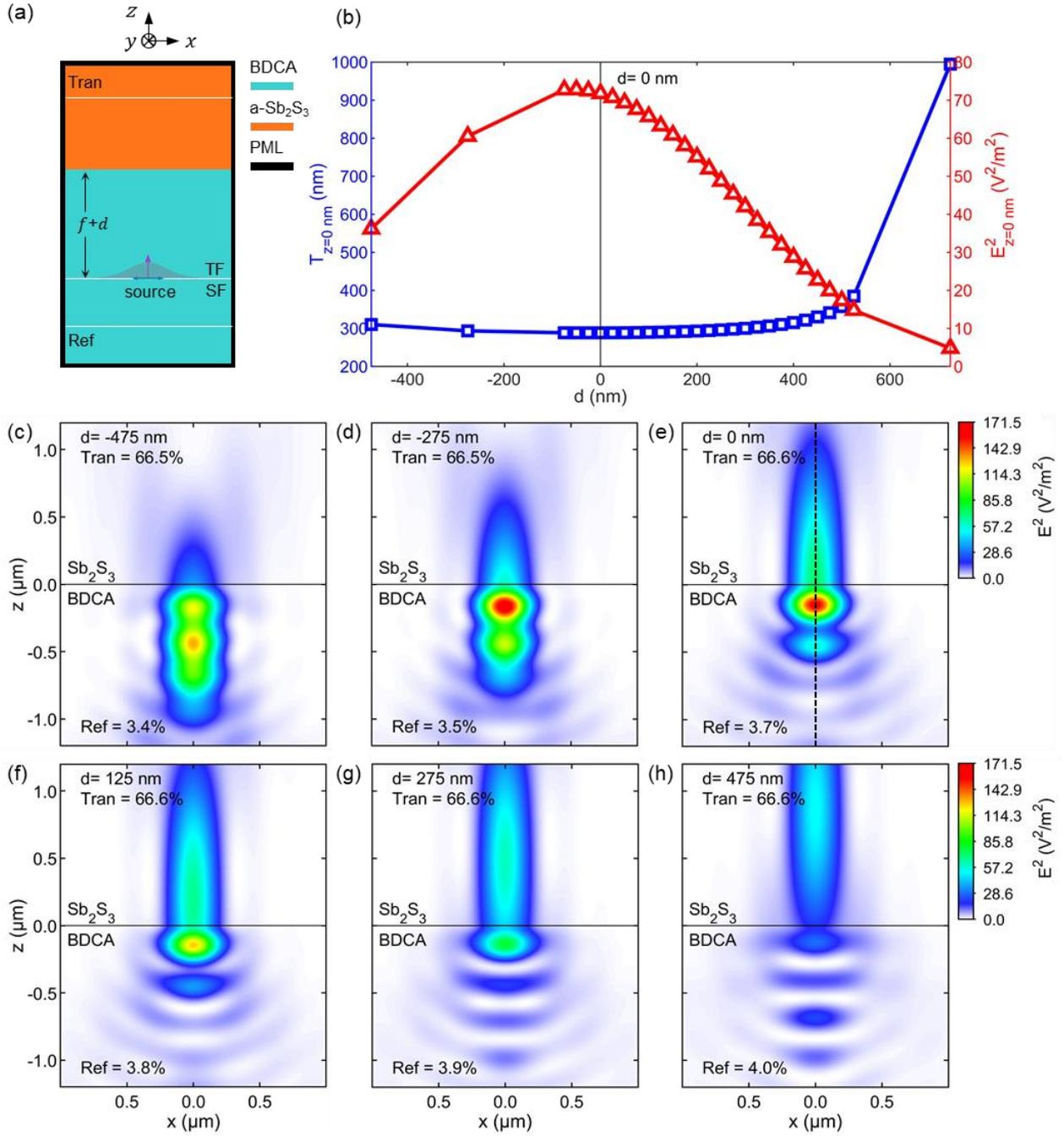

**Figure S4.** (a) xz-plane perspective view of the FDTD simulation setup of tightly focused beam in BDCA/a-Sb$_2$S$_3$ interface, $f = -1.925$ μm is the distance below the interface ($z$ =0 nm) and the source plane, $d$ is the offset of the source plane position, $d = 0$ nm denotes in-focus. Negative $d$ values denote the downshift of source plane (over-focus). Postive $d$ values denote the upshift of source plane (under-focus). (b) Blue line: Beam waist at $z = 0$ nm position (noted as $T_{z=0\,nm}$). Red line: Maximum intensity at $z = 0$ nm position (noted as $E^2_{z=0\,nm}$). xz-plane intensity ($E^2 = E_x^2 + E_y^2 + E_z^2$) map when (c) $d = -475$ nm, (d) $-275$ nm, (e) 0 nm, (f) 125 nm, (g) 275 nm, (h) 475 nm.

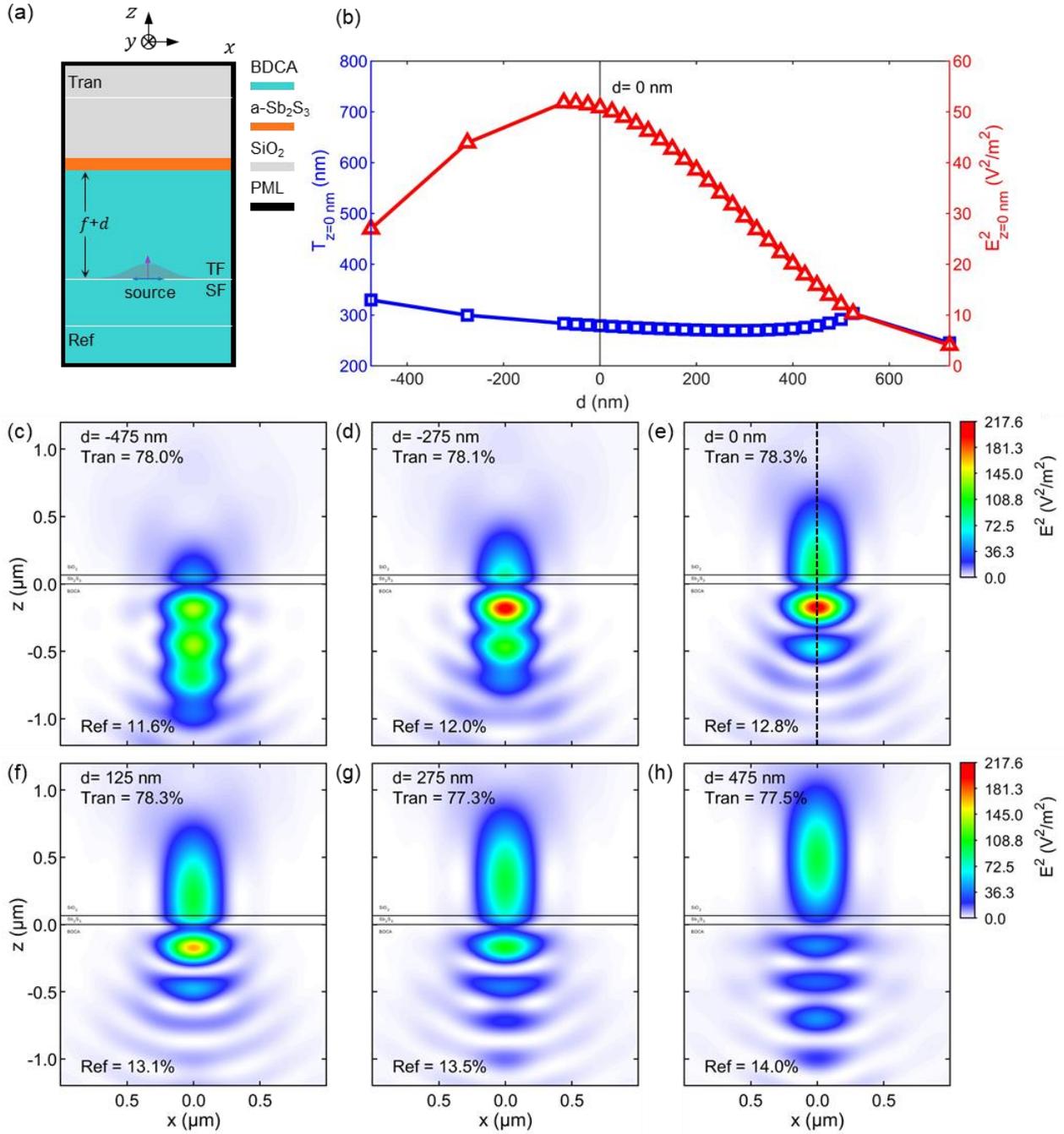

**Figure S5.** (a) xz-plane perspective view of the FDTD simulation setup of tightly focusing in BDCA/a-Sb$_2$S$_3$/SiO$_2$ interface, $f = -1.925$ μm is the distance below the BDCA/a-Sb$_2$S$_3$ interface ($z = 0$ nm) and the source plane, $d$ is the offset of the source plane position. The thickness of a-Sb$_2$S$_3$ is 66 nm. (b) Blue line: Beam waist at $z = 0$ nm position (noted as $T_{z=0\,nm}$). Red line: Maximum intensity at $z = 0$ nm position (noted as $E^2_{z=0\,nm}$). xz-plane intensity ($E^2 = E_x^2 + E_y^2 + E_z^2$) map when (c) $d = -475$ nm, (d) $-275$ nm, (e) 0 nm, (f) 125 nm, (g) 275 nm, (h) 475 nm.

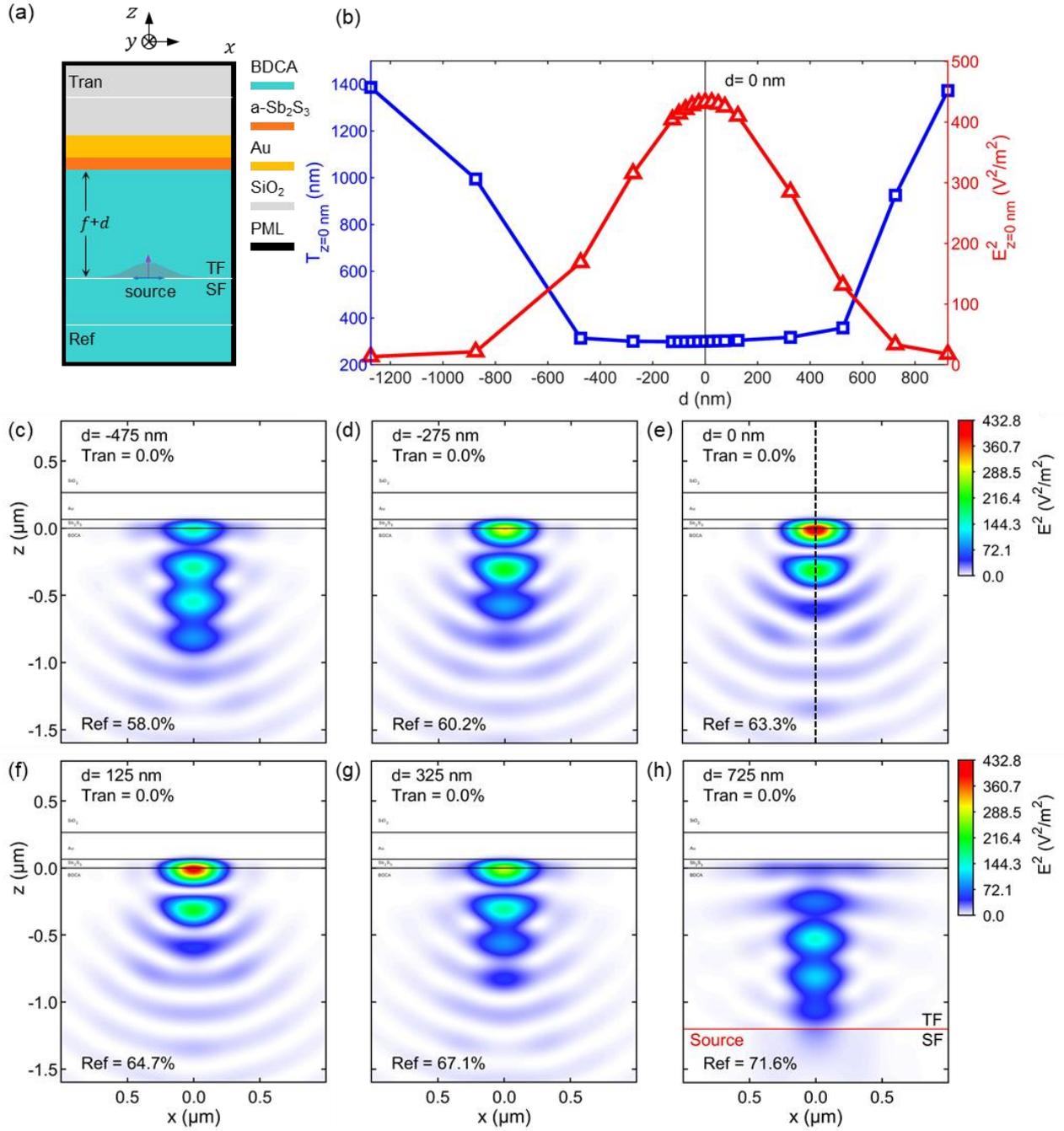

**Figure S6.** (a) xz-plane perspective view of the FDTD simulation setup of tightly focusing in BDCA/a-Sb$_2$S$_3$/Au/SiO$_2$ interface, $f = -1.925$ μm is the distance below the interface ($z = 0$ nm) and the source plane, $d$ is the offset of the source plane position. The thickness of a-Sb$_2$S$_3$ and Au are 66 nm and 200 nm, respectively. (b) Blue line: Beam waist at $z = 0$ nm position (noted as $T_{z=0\,nm}$). Red line: Maximum intensity at $z = 0$ nm position (noted as $E^2_{z=0\,nm}$). xz-plane intensity ($E^2 = E_x^2 + E_y^2 + E_z^2$) map when (c) $d = -475$ nm, (d) $-275$ nm, (e) 0 nm, (f) 125 nm, (g) 325 nm, (h) 725 nm.

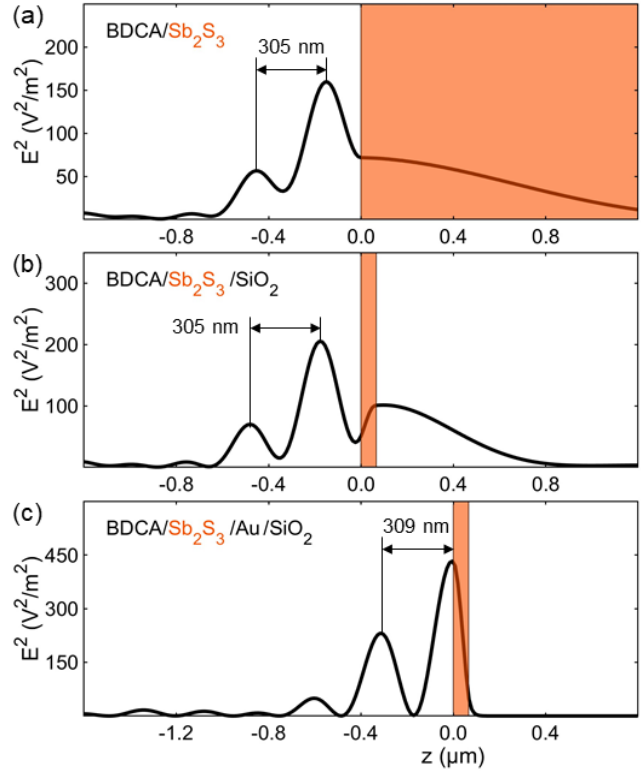

**Figure S7.** The longitudinal intensity ($E^2$) profile along the dashed lines in (a) Fig. S4e, (b) Fig. S5e, and (c) Fig. S6e. (a) For the BDCA/a-Sb$_2$S$_3$ interface, the maximum intensity is 151.8 $V^2/m^2$ at $z = -150$ nm. (b) For the BDCA/a-Sb$_2$S$_3$/SiO$_2$ interface, the maximum intensity is 205.6 $V^2/m^2$ at $z = -177$ nm. (c) For the BDCA/a-Sb$_2$S$_3$/Au/SiO$_2$ interface, the maximum intensity is 432.8 $V^2/m^2$ at $z = 0$ nm. $z = 0$ nm is always set as the interface between BDCA and a-Sb$_2$S$_3$. Once see that the distance between the 1st and 2end peaks in the BDCA is always around 300 nm

## Part SII: Overview of the test patterns

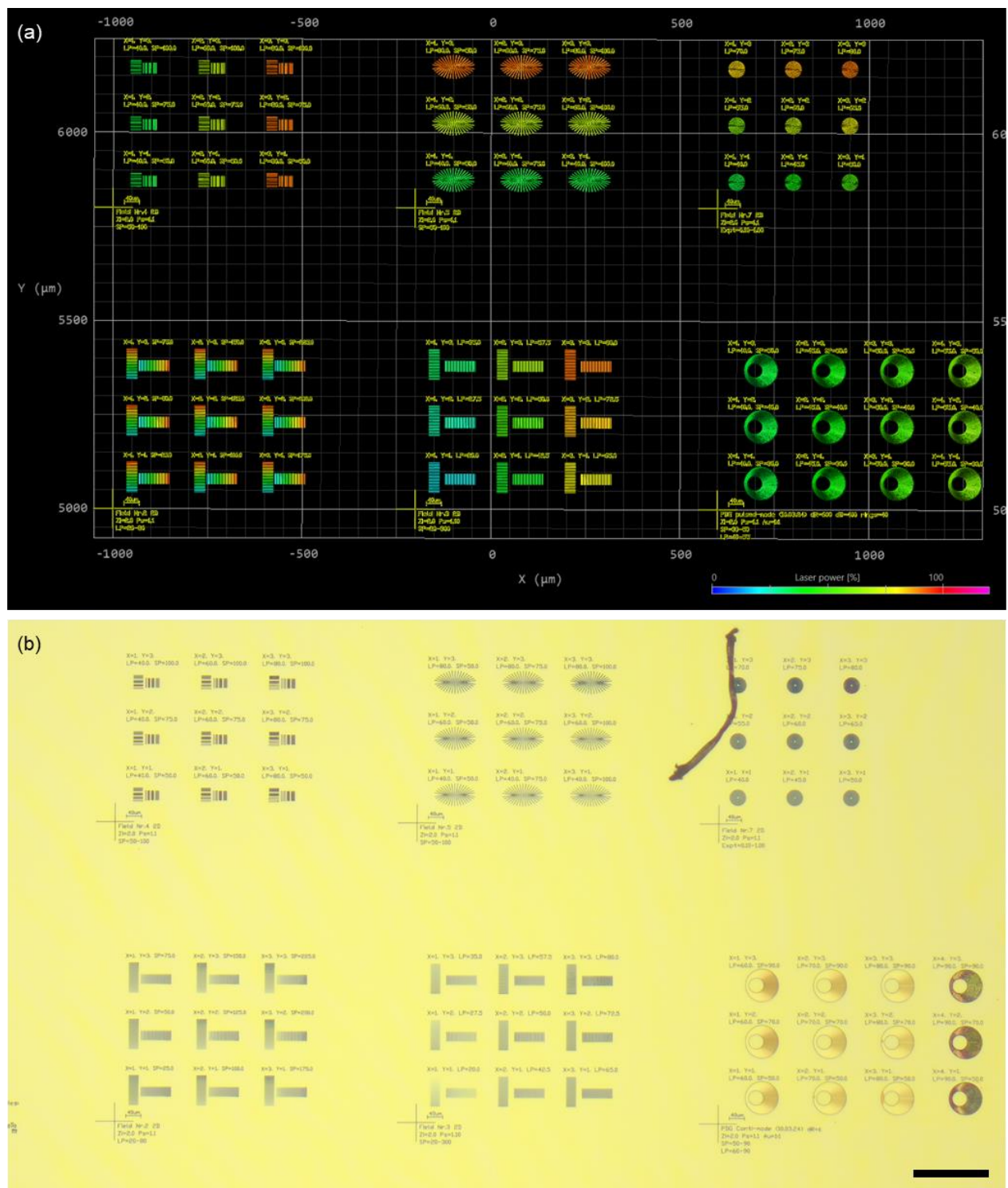

**Figure S8.** (a) The design of the structures presented in this work, rendered with Describe software from Nanoscribe GmbH. All the structures are code defined. (b) Bright field (BF) optical microscope image of the printed structures on a-$Sb_2S_3$ (66nm)/Au (200 nm) film. Scale bar: 200 μm.

**Reference:**

1. Wilson, T.    (Academic Press, 1990).
2. Hell, S. W. & Wichmann, J. Breaking the diffraction resolution limit by stimulated emission: stimulated-emission-depletion fluorescence microscopy. *Optics letters* **19**, 780-782 (1994).
3. Sun, H.-B. *et al.* Experimental investigation of single voxels for laser nanofabrication via two-photon photopolymerization. *Applied physics letters* **83**, 819-821 (2003).
4. Li, Z.-Z. *et al.* Super-stealth dicing of transparent solids with nanometric precision. *Nature Photonics*, 1-10 (2024).
5. Huang, B., Bates, M. & Zhuang, X. Super-resolution fluorescence microscopy. *Annual review of biochemistry* **78**, 993-1016 (2009).
6. Zhao, M. *et al.* A 3D nanoscale optical disk memory with petabit capacity. *Nature* **626**, 772-778 (2024).
7. Grier, D. G. A revolution in optical manipulation. *nature* **424**, 810-816 (2003).
8. Abbe, E. Beiträge zur Theorie des Mikroskops und der mikroskopischen Wahrnehmung. *Archiv für mikroskopische Anatomie* **9**, 413-468 (1873).
9. Hecht, E. *Optics*.  (Pearson Education India, 2012).
10. Abbe, E. *Neue Apparate zur Bestimmung des Brechungs-und Zerstreuungsvermögens fester und flüssiger Körper*.  (Mauke's Verlag, 1874).
11. Li, Q. *Optimization of point spread function of a high numerical aperture objective lens: application to high resolution optical imaging and fabrication*, École normale supérieure de Cachan-ENS Cachan, (2014).
12. Murray, J. M., Wei, J., Barnes, J. O., Slagle, J. E. & Guha, S. Measuring refractive index using the focal displacement method. *Applied optics* **53**, 3748-3752 (2014).
13. Born, M. & Wolf, E. *Principles of optics: electromagnetic theory of propagation, interference and diffraction of light*.  (Elsevier, 2013).
14. Török, P., Varga, P., Laczik, Z. & Booker, G. Electromagnetic diffraction of light focused through a planar interface between materials of mismatched refractive indices: an integral representation. *JOSA A* **12**, 325-332 (1995).
15. Richards, B. & Wolf, E. Electromagnetic diffraction in optical systems, II. Structure of the image field in an aplanatic system. *Proceedings of the Royal Society of London. Series A. Mathematical and Physical Sciences* **253**, 358-379 (1959).
16. Novotny, L. & Hecht, B. *Principles of nano-optics*.  (Cambridge university press, 2012).
17. Rumpf, R. C. *Electromagnetic and Photonic Simulation for the Beginner: Finite-Difference Frequency-Domain in MATLAB®*.  (Artech House, 2022).
18. Noll, R. J. Zernike polynomials and atmospheric turbulence. *JOsA* **66**, 207-211 (1976).
19. Zernike, F. Diffraction theory of the knife-edge test and its improved form, the phase-contrast method. *Monthly Notices of the Royal Astronomical Society, Vol. 94, p. 377-384* **94**, 377-384 (1934).

## 9. References:

1. Jia, W., Menon, R. & Sensale-Rodriguez, B. Visible and near-infrared programmable multi-level diffractive lenses with phase change material Sb 2 S 3. *Optics Express* **30**, 6808-6817 (2022).
2. Dong, W. *et al.* Wide bandgap phase change material tuned visible photonics. *Advanced Functional Materials* **29**, 1806181 (2019).
3. Eggleton, B. J., Luther-Davies, B. & Richardson, K. Chalcogenide photonics. *Nature Photonics* **5**, 141-148 (2011). https://doi.org:10.1038/nphoton.2011.309
4. Zhang, J. *et al.* All-organic polymeric materials with high refractive index and excellent transparency. *Nat Commun* **14**, 3524 (2023). https://doi.org:10.1038/s41467-023-39125-w
5. Biegański, A. *et al.* Sb2S3 as a low-loss phase-change material for mid-IR photonics. *Optical Materials Express* **14** (2024). https://doi.org:10.1364/ome.511923
6. Kim, Y. B., Cho, J. W., Lee, Y. J., Bae, D. & Kim, S. K. High-index-contrast photonic structures: a versatile platform for photon manipulation. *Light Sci Appl* **11**, 316 (2022). https://doi.org:10.1038/s41377-022-01021-1
7. Wang, W. *et al.* Grayscale Electron Beam Lithography Direct Patterned Antimony Sulfide. *arXiv preprint arXiv:2401.13427* (2024).
8. Delaney, M., Zeimpekis, I., Lawson, D., Hewak, D. W. & Muskens, O. L. A new family of ultralow loss reversible phase‐change materials for photonic integrated circuits: Sb2S3 and Sb2Se3. *Advanced functional materials* **30**, 2002447 (2020).
9. Liu, H. *et al.* Rewritable color nanoprints in antimony trisulfide films. *Science advances* **6**, eabb7171 (2020).
10. Chen, R. *et al.* Non-volatile electrically programmable integrated photonics with a 5-bit operation. *Nature Communications* **14**, 3465 (2023).
11. Wang, W., Boneberg, J. & Schmidt-Mende, L. Performance enhancement in Sb2S3 solar cell processed with direct laser interference patterning. *Solar Energy Materials and Solar Cells* **230**, 111235 (2021).
12. Wang, W., Pfeiffer, P. & Schmidt‐Mende, L. Direct patterning of metal chalcogenide semiconductor materials. *Advanced Functional Materials* **30**, 2002685 (2020).
13. Tan, J. W., Wang, G., Li, Y., Yu, Y. & Chen, Q. D. Femtosecond Laser Fabrication of Refractive/Diffractive Micro‐Optical Components on Hard Brittle Materials. *Laser & Photonics Reviews* **17** (2023). https://doi.org:10.1002/lpor.202200692
14. Chen, F. & de Aldana, J. R. V. Optical waveguides in crystalline dielectric materials produced by femtosecond‐laser micromachining. *Laser & Photonics Reviews* **8**, 251-275 (2013). https://doi.org:10.1002/lpor.201300025
15. Miller, F. *et al.* Rewritable Photonic Integrated Circuit Canvas Based on Low-Loss Phase Change Material and Nanosecond Pulsed Lasers. *Nano Letters* (2024).
16. Yu, N. & Capasso, F. Flat optics with designer metasurfaces. *Nat Mater* **13**, 139-150 (2014). https://doi.org:10.1038/nmat3839
17. Choudhury, S. M. *et al.* Material platforms for optical metasurfaces. *Nanophotonics* **7**, 959-987 (2018). https://doi.org:10.1515/nanoph-2017-0130
18. Lin, F.-C. *et al.* Designable spectrometer-free index sensing using plasmonic Doppler gratings. *Analytical chemistry* **91**, 9382-9387 (2019).
19. Nomura, R., Inazawa, S. j., Kanaya, K. & Matsuda, H. Thermal decomposition of butylindium thiolates and preparation of indium sulfide powders. *Applied Organometallic Chemistry* **3**, 195-197 (1989).
20. Wang Wei, Volker Deckert, Jer-Shing Huang & Schmidt, M. Direct 3D printing of high refractive index nanostructures (3D-HiRes). *Leibniz Institute of Photonic Technology (IPHT) Innovation Project* (2021/2022).
21. Wang, X. *et al.* A fast chemical approach towards Sb 2 S 3 film with a large grain size for high-performance planar heterojunction solar cells. *Nanoscale* **9**, 3386-3390 (2017).
22. Abbe, E. Beiträge zur Theorie des Mikroskops und der mikroskopischen Wahrnehmung. *Archiv für mikroskopische Anatomie* **9**, 413-468 (1873).
23. Cao, H.-Z. *et al.* Two-photon nanolithography of positive photoresist thin film with ultrafast laser direct writing. *Applied Physics Letters* **102** (2013).

24   Heiskanen, S., Geng, Z., Mastomäki, J. & Maasilta, I. J. Nanofabrication on 2D and 3D topography via positive‐tone direct‐write laser lithography. *Advanced Engineering Materials* **22**, 1901290 (2020).
25   Mack, C. A. Analytical expression for the standing wave intensity in photoresist. *Applied optics* **25**, 1958-1961 (1986).
26   Abbe, E. *Neue Apparate zur Bestimmung des Brechungs-und Zerstreuungsvermögens fester und flüssiger Körper*.  (Mauke's Verlag, 1874).
27   Grushina, A. Direct-write grayscale lithography. *Advanced Optical Technologies* **8**, 163-169 (2019).
28   Yermakov, O. *et al.* Advanced fiber in-coupling through nanoprinted axially symmetric structures. *Applied Physics Reviews* **10** (2023). https://doi.org:10.1063/5.0127370
29   Zeisberger, M., Schneidewind, H., Wieduwilt, T., Yermakov, O. & Schmidt, M. A. Nanoprinted microstructure-assisted light incoupling into high-numerical aperture multimode fibers. *Opt Lett* **49**, 1872-1875 (2024). https://doi.org:10.1364/OL.521471
30   Huang, L. *et al.* Sub-wavelength patterned pulse laser lithography for efficient fabrication of large-area metasurfaces. *Nature communications* **13**, 5823 (2022).
31   Li, Q.-H. *et al.* Two-photon shape-modulated maskless lithography of positive photoresist of S1813. *Optical Materials* **137**, 113509 (2023).
32   Li, Z.-Z. *et al.* Super-stealth dicing of transparent solids with nanometric precision. *Nature Photonics* **18**, 799-808 (2024).
33   Khorasaninejad, M. *et al.* Metalenses at visible wavelengths: Diffraction-limited focusing and subwavelength resolution imaging. *Science* **352**, 1190-1194 (2016).
34   Ren, H. *et al.* An achromatic metafiber for focusing and imaging across the entire telecommunication range. *nature communications* **13**, 4183 (2022).
35   Wang, N., Zeisberger, M., Hübner, U. & Schmidt, M. A. Nanotrimer enhanced optical fiber tips implemented by electron beam lithography. *Optical Materials Express* **8** (2018). https://doi.org:10.1364/ome.8.002246
36   Keum, D. *et al.* Xenos peckii vision inspires an ultrathin digital camera. *Light Sci Appl* **7**, 80 (2018). https://doi.org:10.1038/s41377-018-0081-2
37   Chen, W. T., Zhu, A. Y. & Capasso, F. Flat optics with dispersion-engineered metasurfaces. *Nature Reviews Materials* **5**, 604-620 (2020).
38   Shrestha, S., Overvig, A. C., Lu, M., Stein, A. & Yu, N. Broadband achromatic dielectric metalenses. *Light Sci Appl* **7**, 85 (2018). https://doi.org:10.1038/s41377-018-0078-x
39   Vijayakumar, A. & Bhattacharya, S.    (SPIE).
40   Wang, Z. *et al.* High efficiency and scalable fabrication of fresnel zone plates using holographic femtosecond pulses. *Nanophotonics* **11**, 3081-3091 (2022).
41   Chen, Y., Ding, Y., Yu, H. & Li, X. in *Photonics.*  249 (MDPI).
42   Li, Z. *et al.* Generation of an Ultra‐Long Transverse Optical Needle Focus Using a Monolayer MoS2 Based Metalens. *Advanced Optical Materials*, 2402024
43   Wu, Y. & Ozcan, A. Lensless digital holographic microscopy and its applications in biomedicine and environmental monitoring. *Methods* **136**, 4-16 (2018). https://doi.org:10.1016/j.ymeth.2017.08.013
44   Wilson, T.   (Academic Press, 1990).
45   Hell, S. W. & Wichmann, J. Breaking the diffraction resolution limit by stimulated emission: stimulated-emission-depletion fluorescence microscopy. *Optics letters* **19**, 780-782 (1994).
46   Sun, H.-B. *et al.* Experimental investigation of single voxels for laser nanofabrication via two-photon photopolymerization. *Applied physics letters* **83**, 819-821 (2003).
47   Li, Z.-Z. *et al.* Super-stealth dicing of transparent solids with nanometric precision. *Nature Photonics*, 1-10 (2024).
48   Huang, B., Bates, M. & Zhuang, X. Super-resolution fluorescence microscopy. *Annual review of biochemistry* **78**, 993-1016 (2009).
49   Zhao, M. *et al.* A 3D nanoscale optical disk memory with petabit capacity. *Nature* **626**, 772-778 (2024).
50   Grier, D. G. A revolution in optical manipulation. *nature* **424**, 810-816 (2003).
51   Hecht, E. *Optics.*  (Pearson Education India, 2012).

52   Li, Q. *Optimization of point spread function of a high numerical aperture objective lens: application to high resolution optical imaging and fabrication*, École normale supérieure de Cachan-ENS Cachan, (2014).
53   Murray, J. M., Wei, J., Barnes, J. O., Slagle, J. E. & Guha, S. Measuring refractive index using the focal displacement method. *Applied optics* **53**, 3748-3752 (2014).
54   Born, M. & Wolf, E. *Principles of optics: electromagnetic theory of propagation, interference and diffraction of light*. (Elsevier, 2013).
55   Török, P., Varga, P., Laczik, Z. & Booker, G. Electromagnetic diffraction of light focused through a planar interface between materials of mismatched refractive indices: an integral representation. *JOSA A* **12**, 325-332 (1995).
56   Richards, B. & Wolf, E. Electromagnetic diffraction in optical systems, II. Structure of the image field in an aplanatic system. *Proceedings of the Royal Society of London. Series A. Mathematical and Physical Sciences* **253**, 358-379 (1959).
57   Novotny, L. & Hecht, B. *Principles of nano-optics*. (Cambridge university press, 2012).
58   Rumpf, R. C. *Electromagnetic and Photonic Simulation for the Beginner: Finite-Difference Frequency-Domain in MATLAB®*. (Artech House, 2022).
59   Noll, R. J. Zernike polynomials and atmospheric turbulence. *JOsA* **66**, 207-211 (1976).
60   Zernike, F. Diffraction theory of the knife-edge test and its improved form, the phase-contrast method. *Monthly Notices of the Royal Astronomical Society, Vol. 94, p. 377-384* **94**, 377-384 (1934).